\documentclass[twocolumn,showpacs,prl,preprintnumbers,superscriptaddress,amsmath,amssymb]{revtex4}

\usepackage{graphicx}
\usepackage{psfig}
\usepackage{picins}

\newcommand{\grad}{$^\circ$}
\newcommand{\fcs}{FeCr$_2$S$_4$}

\begin{document}

 
\title{Low-temperature structural transition in FeCr$_2$S$_4$}

\author{M. Mertinat}
\affiliation{%
Institut f\"ur Physik, Universit\"at Augsburg,
D-86135 Augsburg, Germany}
\email{Markus.Mertinat@Physik.Uni-Augsburg.DE}
\author{V. Tsurkan}
\affiliation{%
Institut f\"ur Physik, Universit\"at Augsburg,
D-86135 Augsburg, Germany}
\affiliation{%
Institute of Applied Physics, Academy of Sciences of Moldova,
MD 2028 Chisinau, R. Moldova}
\author{D. Samusi}
\affiliation{%
Institute of Applied Physics, Academy of Sciences of Moldova,
MD 2028 Chisinau, R. Moldova}
\author{R. Tidecks}
\author{F. Haider}
\affiliation{%
Institut f\"ur Physik, Universit\"at Augsburg,
D-86135 Augsburg, Germany}

\date{\today}
\begin{abstract}
\noindent Transmission electron microscopy studies of [110]
and [111] oriented FeCr$_2$S$_4$ single crystals at different
temperatures reveal a structural transition at low temperatures
indicating a cubic-to-triclinic symmetry reduction within
crystallographic domains.
The overall crystal symmetry was found to be reduced from
Fd3m to F${\bar 4}$3m.
The triclinic distortions were suggested to result from the combined
actions of tetragonal distortions due to the Jahn-Teller active 
Fe$^{2+}$ ions and trigonal distortions due to a displacement of the
Cr$^{3+}$ ions in the $\left<111\right>$ direction.
\end{abstract}


\maketitle

Spin-lattice coupling in highly correlated magnetic systems plays
an important role in the formation of the magnetic ground state and 
governs the electronic properties.
Electron-phonon interaction and lattice polarons
contribute essentially to the colossal magnetoresistance (CMR)
effect and the magnetic field induced metal-insulator transition
in manganite perovskites \cite{fcs:Millis:95,fcs:Millis:96}.
In addition to small lattice polarons, correlated lattice 
distortions may appear and result in nanoscale regions
with charge and orbital order \cite{fcs:Dai:00}.
These regions are connected with the change of the structural
symmetry and are also responsible for the magnetotransport
anomalies.

In this letter we report the observation of a structural
lattice transformation at low temperatures in another CMR material,
ternary ferrimagnetic \fcs\ with a cubic spinel-type
crystal structure \cite{fcs:Ramirez:97}.
The existence of local structural distortions and the possibility
of a structural transformation in this compound was suggested already in 1967 based
on the appearance of quadrupole splitting and a low-temperature
anomaly of the electric field gradient induced on the tetrahedrally
coordinated Fe$^{2+}$ ions \cite{fcs:Eibschuetz:67,fcs:Spender:72a}.
These features were attributed to a strong coupling between the 
Jahn-Teller (J-T) active ferrous ions which allows a distortion of the 
FeS$_4$ tetrahedrons \cite{fcs:Feiner:82} and were explained in the framework of
static and dynamic J-T effects.
An alternative explanation suggested another type of orbital ordering
due to a hybridization of Cr and Fe states \cite{fcs:Brossard:79}.
The interpretation of the M{\"o}{\ss}bauer data, however, was in conflict with
X-ray and neutron scattering diffraction investigations, which
found that polycrystalline \fcs\ remains a cubic spinel
down to 4.2 K \cite{fcs:Shirane:64,fcs:Broquetas:64}.
In powdered single crystals, the symmetry was found to be
unchanged, too, although a broadening of the X-ray diffraction
lines was observed and was suggested as due to inhomogeneous
lattice distortions below the Curie temperature down to 4.2~K
\cite{fcs:Goebel:76}.

Several recent experimental investigations on \fcs\ single crystals pointed
out the importance of a spin-lattice coupling. 
A cusp-like anomaly
in the temperature dependence of the magnetization at $T_m\approx60$~K 
and a splitting of zero-field cooled and field cooled magnetization below this
temperature was observed \cite{fcs:Tsurkan:01c}.
Hydrostatic pressure
investigations \cite{fcs:Tsurkan:01b} show that the magnetic
anomaly at $T_m$ in \fcs\ is strongly sensitive to lattice contraction.
$T_m$ is increased by pressure 
with a rate of d$T_m$/d$p\approx 30$~K/GPa. This behavior was related to 
the appearance of a non-cubic magnetocrystalline anisotropy due to
structural distortions.
AC-susceptibility measurements \cite{fcs:Tsurkan:01a}
and magnetoresistance studies \cite{fcs:Tsurkan:01b} indicate
that the spin-glass-like features below $T_m$ 
are connected with changes in the domain structure due to
additional pinning centers which were suggested to appear as a
result of a structural lattice transformation.
Very recently ultrasonic measurements of \fcs\ single crystals
gave evidence for a structural transformation at $\approx 60$~K.
The elastic moduli manifest a step-like feature around this temperature
indicating a structural phase transition of first-order type.
Below 60~K a pronounced softening of the elastic moduli was found.
The experimental data indicate the appearance of a trigonal
distortion which was explained in terms of an orbital ordering
with coupling of the orbitals of Fe ions along the
$\left<111\right>$ direction \cite{fcs:Maurer:03}.

To investigate the structural distortions we used
selected area electron diffraction (SAED) in a transmission electron
microscope (TEM).
It shows a cut through the reciprocal lattice
and, therefore, is very sensitive to small changes of the
local crystal symmetry of the lattice
\cite{Reimer:93}.
Any changes in the crystal symmetry and the lattice
constants can thus be directly observed.

Ternary polycrystalline material
obtained by solid state reaction from high purity elements
was the source to grow \fcs\ single crystals by chemical transport
reaction \cite{Schaefer:62}.
Chromium chloride and tellurium chloride were used as the source
of the transport agent, Cl$_2$. The growth temperature was varied between
820 and 850\grad C.
The resulting crystals grew to a size up to 4-6~mm.
The single phase
spinel structure was confirmed by X-ray diffraction analysis.
Sample composition was checked by electron probe microanalysis,
which revealed a nearly stochiometric composition
(within 2~mole\%) and a small 
amount of chlorine ($\lesssim 0.5$~mole\%).
Plates cut from single crystals in different directions
were glued on a Cu supporting
disc with a 800~$\mu$m diameter centered hole. 
After mechanical dimpling, the samples were thinned
using an Ar$^+$ ion polishing system.
The TEM (Philips CM 12 operating at 120~kV) was
assembled with a low temperature cooling stage.
The double tilt sample holder, Gatan 636.LHe, was 
cooled with liquid helium to a lowest
temperature of 14~K. The diffraction images were recorded
{\em in situ\/} for different temperatures.
The microscope parameters were kept unchanged for a series
of diffraction images in order to minimize possible errors due to
hysteresis effects of the magnetic electron lenses.
The accuracy of the temperature measurement was
better than $\pm3$~K.
The images were recorded with a CCD camera or negative
film plates which were digitized for further processing.

\begin{figure}[htb]
\hpic{
 \psfig{width=1.0\linewidth,figure=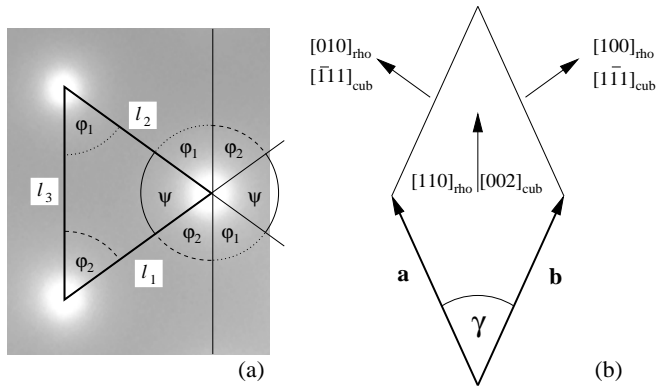}
}
\caption{\label{fig:rhombus} (a) SAED pattern of
[110] zone axis orientation with the triangle used for evaluation
of distances and angles;
(b) [001] view of rhombohedral primitive cell with the cubic and 
rhombohedral directions.}
\end{figure}

To evaluate the SAED patterns and determine the
spot positions a special computer algorithm was developed \cite{Mertinat:03}.
The distances and angles between neighboring spots were calculated
and averaged.
The distances in the diffraction patterns, $l_i$, were transformed into
interplanar spacings in direct space, $d_i$.
Relative changes less than $10^{-4}$ can be detected.
To construct the unit cell, samples with two different
orientations ([110] and [111]) are needed.
In the cubic structure, for the [110] orientation
the diffraction pattern should exhibit two equal angles
(noted $\varphi_1 = \varphi_2 \approx 54.7^\circ$ in
Fig.~\ref{fig:rhombus}a) and a third one,
$\psi\approx 70.6$\grad.
From the three distances between the spots two must be equal
(noted as $l_1$ and $l_2$ in Fig.~\ref{fig:rhombus}a) and
differ from the third one
($l_3$). The ratio $l_1$/$l_3$ = $l_2$/$l_3$ $\approx 1/1.15$.
In the diffraction pattern of an ideal fcc crystal in [111] direction
all distances between the neighbored spots
are equal and all angles between these spots are 60\grad
\cite{Reimer:93}.\\

\begin{figure}[htb]
\hpic{
 \includegraphics[width=\linewidth]{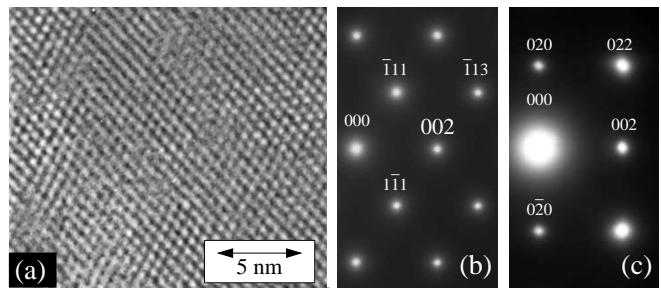}
}
{\caption{\label{fcs:hrtem}
  (a) High resolution image of [110]--\fcs. (b) SAED pattern of
  [110] and (c) [100] oriented samples.
}}
\end{figure}

Figures~\ref{fcs:hrtem}a and b show a high resolution image and SAED pattern
of the \fcs\ single crystal in [110] zone axis orientation.
The image indicates a quite regular
lattice structure without defects such as precipitates
and dislocations (Fig.~\ref{fcs:hrtem}a).
The corresponding diffraction pattern shows
the ordinary reflections (000) and (111), but also the (002)
reflection which is forbidden for the Fd3m space group (Fig.~\ref{fcs:hrtem}b).
It was observed in all sample regions with different thicknesses
at all temperatures.
To verify whether this is an intrinsic reflection or is caused by
double diffraction we performed TEM investigations on a sample with
[100] orientation.
For this orientation no contribution from double diffraction is allowed.
Indeed, for [100] oriented samples we also observed small but clearly 
pronounced (200) reflections (Fig.~\ref{fcs:hrtem}c).
Our results are similar to that observed
in MgAl$_2$O$_4$ spinel with reduced symmetry \cite{fcs:Hwang:73} and
suggest that the space group of \fcs\ is F$\bar 4$3m rather than Fd3m.

\begin{figure}[htb]
\hpic[][b]{
 \includegraphics[width=\linewidth]{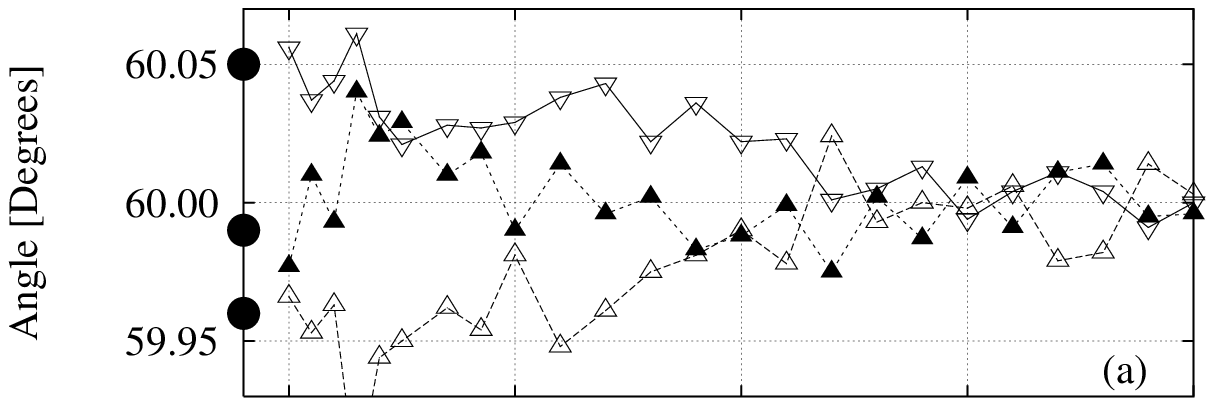}
}
\hpic[][t]{
 \includegraphics[width=\linewidth]{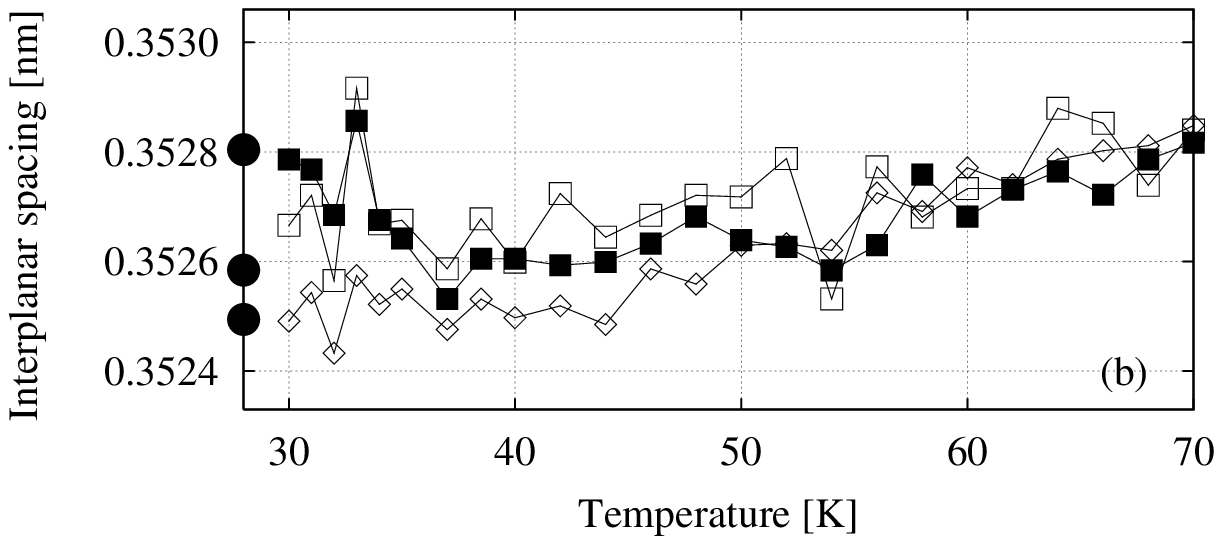}
}
\hpic[][b]{
 \includegraphics[width=\linewidth]{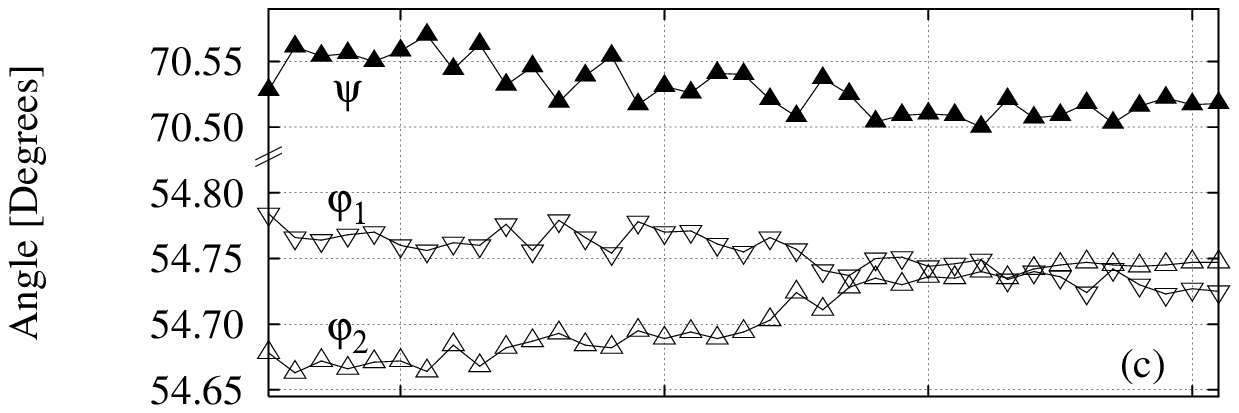}
}
\hpic[][t]{
 \includegraphics[width=\linewidth]{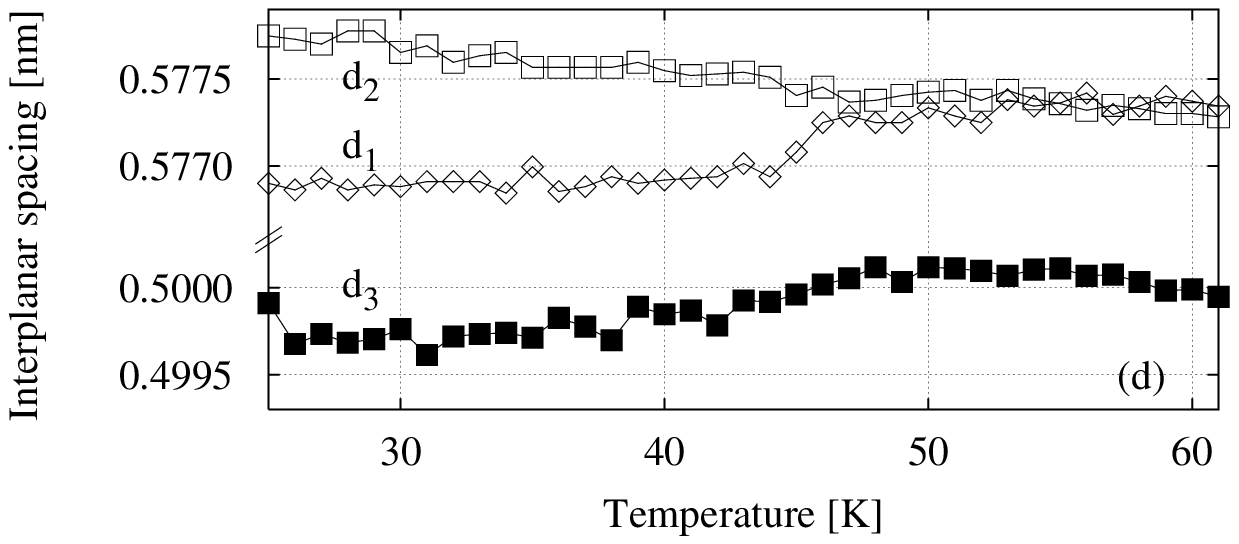}
}
{\caption{\label{fcs:wk_di}
  Angle and interplanar spacing of the diffraction patterns
  as function of temperature
  of [111]-- (a and b) and [110]-- (c and d) oriented \fcs.
  Full dots : Calculated values for the low-temperature structure.
}}
\end{figure}

\begin{figure}[htb]
\hpic[][b]{
 \includegraphics[height=\linewidth,angle=270]{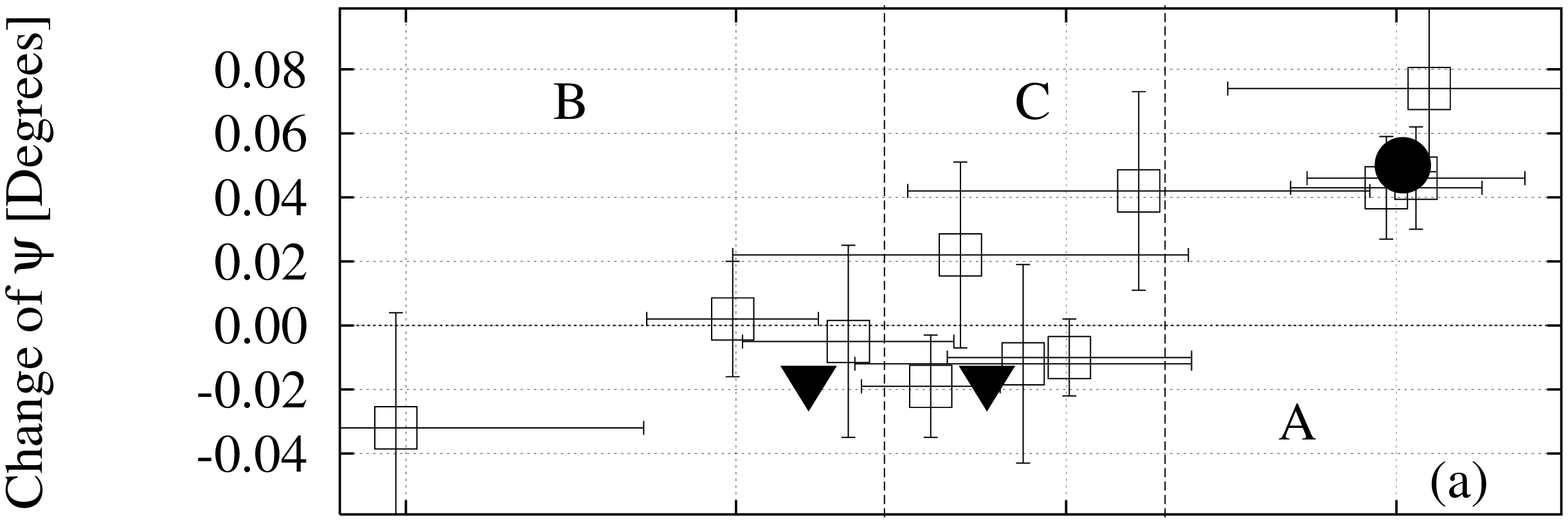}
}
\hpic[][t]{
 \includegraphics[height=\linewidth,angle=270]{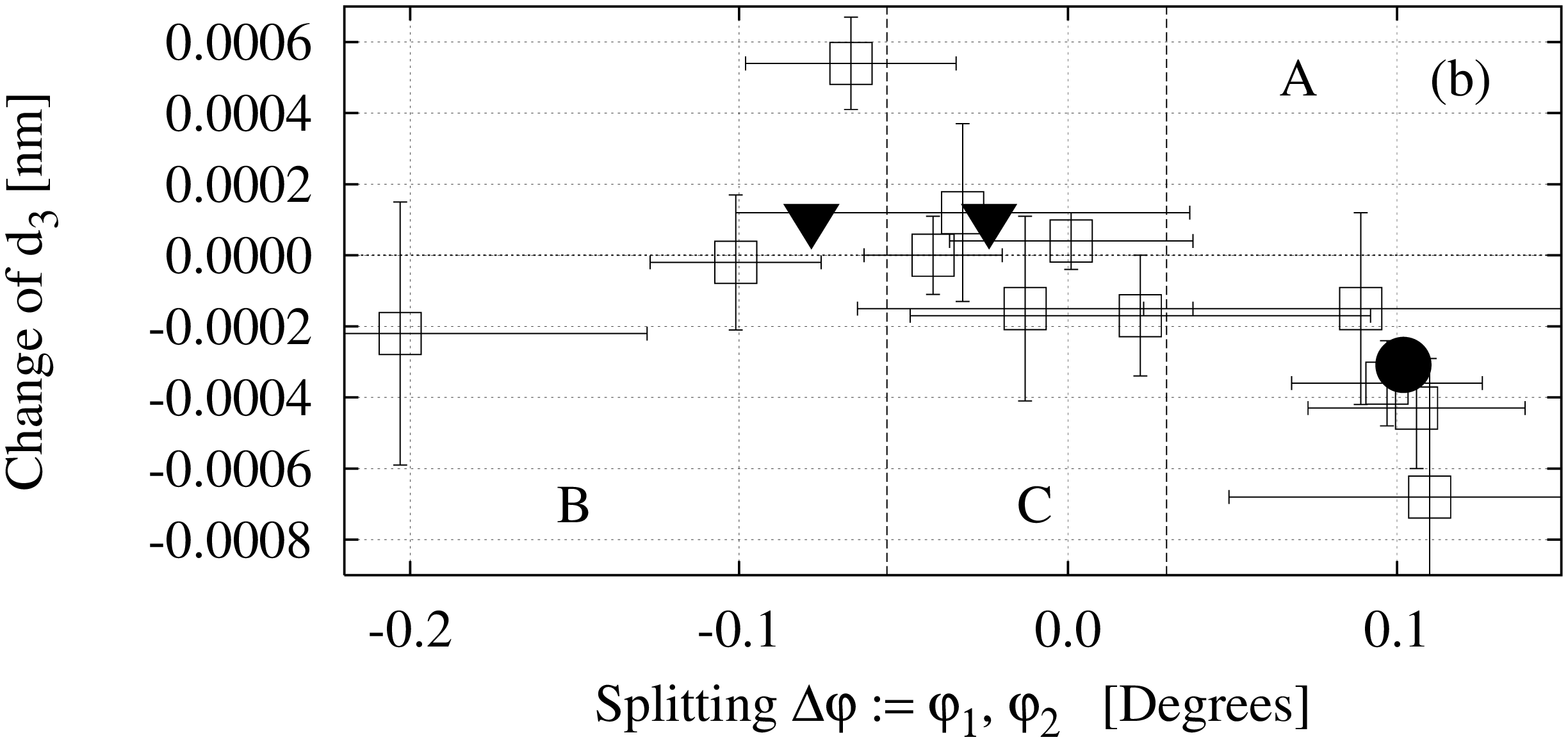}
}
{\caption{\label{fcs:res_figure}
  Variation of the diffraction pattern parameters 
  of [110]--\fcs. Symbols: See text.
}}
\end{figure}

In Fig.~\ref{fcs:wk_di} the temperature dependence of the 
diffraction pattern parameters are shown for the [111] oriented sample. 
Above 60~K the angles and spacings are equal, as expected for a fcc crystal.
Below, a noticeable
difference between the angles appears (Fig.~\ref{fcs:wk_di}a), whereas the
interplanar spacings (Fig.~\ref{fcs:wk_di}b) show a less pronounced
deviation.
For the [110] orientation we observed three different types of behavior
of the diffraction pattern parameters for different sample positions,
noted below as types A, B and C. For the A-type patterns
the parameters are shown in
Fig.~\ref{fcs:wk_di}c and d.
Below about 50~K clear
differences between angles $\varphi_1$ and $\varphi_2$ (Fig.~\ref{fcs:wk_di}c)
and spacings $d_1$ and $d_2$ (Fig.~\ref{fcs:wk_di}d) are found.
The values of $\varphi_1$  and $d_2$
increase, whereas $\varphi_2$ and $d_1$ decrease for decreasing temperature.
The third parameters
($\psi$ and $d_3$) also show a clear increase and decrease, respectively.
The B-type pattern is also characterised by a splitting of
$\varphi_1$ and $\varphi_2$, and $d_1$ and $d_2$ below 50~K \cite{annot_60K},
but with an opposite and less pronounced variation 
compared to type A.
For the C-type behaviour no
splitting or other significant changes of the diffraction parameters
were observed.
Figure~\ref{fcs:res_figure} and Tab.~1 summarize the experimental observations of these
three different types of behavior. The variation of $\psi$ and $d_3$
versus the splitting between $\varphi_1$ and $\varphi_2$ 
with temperature is presented in Fig.~\ref{fcs:res_figure}a and b, respectively.    
\begin{table}[h]
\begin{center}
\begin{tabular}{|c||c|c|c|c|}
 \hline
 Type & $\Delta\varphi$ [$^\circ$] & $\Delta\psi$ [$^\circ$] & $\Delta d_{1-2}$ [nm] & $\Delta d_3$ [nm] \\
 \hline \hline
 A      & $+0.102(18)$ & $+0.050(11)$ & $-0.00060(16)$ & $-0.00037(9)$ \\ \hline
 B      & $-0.086(17)$ & $-0.007(15)$ & $+0.00027(23)$ & $+0.00024(7)$ \\ \hline
 C      & $-0.026(17)$ & $-0.009(11)$ & $+0.00005(13)$ & $-0.00000(7)$ \\ \hline
\end{tabular}
\caption
  {Splitting of the diffraction pattern parameters for the different types in [110]--\fcs.}
\label{fcs:Aufspaltungstabelle}
\end{center}
\end{table}

To understand the results, we
computed the diffraction patterns for different symmetry types of the
unit cell.
As a starting point for the fitting procedure a primitive
rhombohedral unit cell was used
with the lattice parameters
$a=b=c= 2^{-1/2}\cdot a_{\rm cub}$ and $\alpha=\beta=\gamma=60^\circ$,
where $a_{\rm cub}$ is the lattice constant of the cubic unit cell
\cite{fcs:Gorter:54}. The arrangement
of the vectors normal to the base planes (as shown in
Fig.~\ref{fig:rhombus}b) directly corresponds to
the triangular configuration of the diffraction pattern.
The symmetry of the
distorted rhombohedral cell is lowered to triclinic. 
The parameters $c$, $\alpha$ and $\beta$ of the
triclinic cell were kept constant in the simulation because
for the [110] zone axis changes in the [110] direction cannot
be detected.
Fitting of the type A diffraction
pattern data (full dots in Fig.~\ref{fcs:res_figure}) results in
$a =0.7075(2)\cdot a_{\rm cub}$; $b =0.7066(2)\cdot a_{\rm cub}$;
$c=0.7071\cdot a_{\rm cub}$;
$\alpha=\beta=60^\circ$; $\gamma=60.04(1)^\circ$.
Using this crystallographic data the expected changes for the other two
types of patterns were calculated supposing that they are related to
the other two cubic $\left<110\right>$ orientations. The results for that case
are marked by full triangles in Fig.~\ref{fcs:res_figure} and
reproduce quite well the experimental data.
In the same way, we calculated the expected variations for the 
[111] zone axis orientation (full dots in Fig.~\ref{fcs:wk_di}a and b) and obtained
a very reasonable agreement with the experimental data.
This consistence suggests that the three different
types of diffraction patterns observed are connected with the creation of 
different structural domains or twinning planes.
The orientation of these spacial domains alternate
between the three $\left<110\right>$
directions which were equivalent before transition.
The size of these domains was estimated from the lateral sample drift
during the experiment to lie between 15 and
50~$\mu$m.

To get the difference in the diffraction patterns for the cubic
and the triclinic cell we calculated the powder
diffraction patterns using the {\em EMS (electron microscopy image simulation)\/}
program by P.~Stadelmann \cite{um:Stadelmann:87}, yielding changes on
the order of $5\cdot 10^{-4}$.
Such small changes
can hardly be detected by ordinary X-ray or neutron diffraction
where for polycrystalline samples a broadening of the peaks rather
than a splitting is observed
\cite{fcs:Shirane:64,fcs:Broquetas:64,fcs:Goebel:76}.

Concerning the nature of the structural anomaly in \fcs,
several mechanisms may be considered.
Earlier theoretical
\cite{fcs:Goodenough:55,fcs:Dunitz:57,fcs:Wojtowicz:59,fcs:Goodenough:64}
and experimental studies \cite{fcs:Francombe:57,fcs:Arnott:64,fcs:Wold:63}
of a structural transition in transition metal oxide spinels
attributed them mainly to the Jahn-Teller (JT) active Fe$^{2+}$ ions
which stabilize the tetragonal phase. 
In addition the octahedral Cr$^{3+}$ ions may contribute to the
elastic anisotropy.
Depending on the relative strength of the competing effects, a resulting
low-temperature tetragonal or orthorhombic structure is established
\cite{fcs:Goodenough:64,Krupicka:73}.
None of these structures, however, fits our experimental data
\cite{Mertinat:03}.

From our observations of the reduction of the
overall symmetry from Fd3m to F$\bar 4$3m an additional distortion of the
octahedral sites can be inferred.
A similar reduction of the symmetry
was observed in several spinel compounds \cite{fcs:Grimes:72,fcs:Wold:63}
and was attributed to a displacement of the octahedral-site cations
from the center of the octahedron along the $\left<111\right>$ direction
producing a trigonal distortion.
The presence of trigonal distortions in \fcs\ was also revealed by 
ultrasonic experiments \cite{fcs:Maurer:03}.
Therefore, the observed triclinic distortions in our
\fcs\ crystal may result from the combined action of a 
tetragonal distortion due to tetrahedral-site JT Fe$^{2+}$ ions and a
trigonal one due to octahedral-site Cr$^{3+}$ ions.
Since the structural transformation in \fcs\ occurs below the 
cusp-like anomaly in the magnetization at $T_m$, and
keeping in mind the strong magnetocrystalline
anisotropy in \fcs\ \cite{fcs:Stapele:71,fcs:Goldstein:76} 
there is probably a contribution from spin-orbital coupling to the
lattice distortions.
In fact, recent specific heat \cite{Tsurkan:04} and ultrasonic 
studies \cite{fcs:Maurer:03} attributed the low-temperature anomalies
in this compound to the orbital degrees of freedom.
The structural transformation may result from an orbital ordering.
However, the coupling of the orbitals appears to be more 
complicated than the simple trigonal or tetragonal arrangements suggested
in earlier studies \cite{fcs:Tsurkan:01b,fcs:Maurer:03,annot_stochiometry}.

In conclusion, we investigated the
crystal structure of the \fcs\ magnetic spinels by SAED.
We found a structural anomaly below 60~K which we interprete in terms
of a triclinic distortion within crystallographic domains.
In addition we found a (200) reflection which indicates that the crystal
symmetry belongs to the F${\bar 4}$3m symmetry group.
Thus, our study clarifies the long
standing problem of the structural anomaly in \fcs.

\bibliography{bib/fecrs,bib/book2,bib/ultra,bib/my,bib/annot}

\begin{thebibliography}{35}
\bibitem[{ann({\natexlab{a}})}]{annot_60K}
\bibinfo{note}{The difference in the detected transition temperature for
  samples with [110] and [111] orientations may be explained as due to heating
  of the sample by the electron beam and temperature gradient.}
\end{thebibliography}

\end{document}